\documentclass[prb,twoside,twocolumn,showpacs,superscriptaddress,floatfix]{revtex4}
\usepackage{graphicx}
\usepackage{amsmath}
\usepackage{amssymb}

\begin{document}

\title{Comment on ``Superconducting anisotropy and evidence for intrinsic pinning in single crystalline
MgB$_2$''}
\author{M. Angst}
 \email[Email: ]{angst@phys.ethz.ch}
\affiliation{Solid State Physics Laboratory ETH, 8093 Z\"urich,
Switzerland}
\author{R. Puzniak}
\author{A. Wisniewski}
\affiliation{Institute of Physics, Polish Academy of Sciences, Al.
Lotnikow 32/46, 02-668 Warsaw, Poland}
\author{J. Roos}
\author{H. Keller}
\affiliation{Physik-Institut, Universit\"at Z\"urich, 8057
Z\"urich, Switzerland}
\author{J. Karpinski}
\affiliation{Solid State Physics Laboratory ETH, 8093 Z\"urich,
Switzerland}
\date{\today}
\begin{abstract}
In a recent paper [Phys.\ Rev.\ B {\bf 66}, 012501 (2002)], torque
data measured on a MgB$_2$ single crystal in fields from 10 to
$60\,{\text{kOe}}$ at $10\,{\text{K}}$ were presented. The authors
obtained the anisotropy $\gamma$ by fitting a theoretical
expression to the data and concluded that the anisotropy is field
independent $\gamma \approx 4.3$. They also reported the
observation of ``intrinsic pinning'', which they take as
experimental evidence for the occurence of superconductivity in
the boron layers. In this comment, we discuss the range of
validity of the theoretical expression used by the authors and
show that the conclusion of a field independent anisotropy does
not hold. Furthermore, we present torque data measured on two
crystals of MgB$_2$, establishing the extrinsic nature of the peak
in the irreversible torque observed in some crystals.
\end{abstract}
 \maketitle

%\section{Introduction}

In a recent report,\cite{Takahashi02} Takahashi {\em et al.}
presented torque measurements performed on a single crystal of
MgB$_2$. The authors concluded that the anisotropy $\gamma\approx
4.3$ is field independent at $10\,{\text{K}}$, in contrast to an
earlier finding\cite{Angst02MgB2anis} of a pronounced field
dependence of $\gamma$ at higher temperatures. The main purpose of
this comment is to show that the conclusion in Ref.\
\onlinecite{Takahashi02} of a field independent $\gamma$ at
$10\,{\text{K}}$ does not hold, since the expression used in the
analysis is not applicable to all of the data.

In Ref.\ \onlinecite{Takahashi02}, the experimental data for
angular torque dependences in different fields were described by a
formula developed by Kogan,\cite{Kogan88b} which in CGS units has
the following form:
\begin{equation}
\tau = - \frac{ \Phi_{\circ}H V}{64 \pi^2 \lambda_{ab}^2} \left (
1 - \frac{1}{\gamma^2} \right ) \frac{\sin 2
\theta}{\epsilon(\theta)} \ln \left ( {\frac{\eta
H_{c2}^{\|c}}{\epsilon (\theta) H}} \right ), \label{tau_rev}
\end{equation}
where $\epsilon (\theta) = ( \cos^2 \theta + \sin^2 \theta /
\gamma^2 )^{1/2}$, $\theta$ is the angle between the applied
magnetic field and the $c$-axis of the crystal,
$\gamma=(m_c^*/m_{ab}^*)^{1/2}$ is the effective mass anisotropy,
$H_{c2}^{\|c}$ is the upper critical field parallel to the
$c$-axis, $\lambda_{ab}$ is the in-plane penetration depth, $V$ is
the volume of the crystal, $\Phi_{\circ}$ is the flux quantum, and
$\eta$ is a constant of the order of unity depending on the vortex
lattice structure. Equation (\ref{tau_rev}) for the angular
dependence of the reversible torque was derived on the basis of
the anisotropic London model, valid in the limits of fields
$H_{c1}\ll H \ll H_{c2}$, the latter condition is important to
exclude vortex core overlap.

In the analysis, Takahashi {\em et al.} did not fit to the
experimental data all of the parameters of Eq.\ (\ref{tau_rev}),
but assumed the value of $\eta H_{c2}^{\|c}$ to be fixed and equal
to $60\,{\text{kOe}}$, because earlier measurements on a MgB$_2$
crystal from the same source indicated an upper critical field
parallel to the $c$-axis located in this region at low
temperatures.\cite{Xu01b} It should be mentioned, however, that
these $H_{c2}$ values were determined from resistivity
measurements. Later bulk measurements on single
crystals\cite{Angst02MgB2anis,Sologubenko02,Zehetmayer02,Welp02,Lyard02}
found a lower $H_{c2}^{\|c}$ value in the region of
$30-35\,{\text{kOe}}$, in contrast to the resistivity results,
which are affected by surface effects.\cite{Sologubenko02,Welp02}
Furthermore, the data presented by Xu {\em et al.} indicate an
anisotropy of the upper critical fields of $2.6$, a much lower
value than the estimation of Takahashi {\em et al.}. A possible
reason of the large discrepancy is an overestimation of
$H_{c2}^{\|c}$ by Xu {\em et al.} due to alignment problems. Since
different assumptions on the value of the parameter $\eta
H_{c2}^{\|c}$ yield different anisotropies $\gamma$ from a fit of
Eq.\ (\ref{tau_rev}), it would be interesting to see what values
of $\gamma$ would result with different assumptions of $\eta
H_{c2}^{\|c}$.

Problematic, especially if $H_{c2}^{\|c}$ is smaller than
$60\,{\text{kOe}}$, is the application of Eq.\ (\ref{tau_rev}) for
data measured in fields up to $60\,{\text{kOe}}$, particularly
since data measured in the whole angular region from $0$ to
$180\,{\text{deg}}$ are used in the fits. Even if $H_{c2}^{\|c}$
is indeed about $60\,{\text{kOe}}$, the condition $H \ll H_{c2}$
for the validity of the London model, and thus of Eq.\
(\ref{tau_rev}) is not fulfilled for all data analyzed. It should
be visible in the data if $H>H_{c2}^{\|c}$, since in this case the
superconducting torque is zero for angles for which
$H>H_{c2}(\theta)$, as can be seen, for example, in Fig.\ 1 of
Ref.\ \onlinecite{Angst02MgB2anis}. In the data presented in Ref.\
\onlinecite{Takahashi02}, the relatively low signal-to-noise ratio
makes a corresponding direct unambiguous conclusion about the
location of $H_{c2}^{\|c}$ difficult. Nevertheless, in our opinion
Fig.\ 2 of Ref.\ \onlinecite{Takahashi02} indicates that both the
data measured in $60$ and in $50\,{\text{kOe}}$ are above $H_{c2}$
for field directions close to the $c$-axis. This may or may not be
the case for the data measured in $40\,{\text{kOe}}$ as well. We
therefore think that Eq.\ (\ref{tau_rev}) is not applicable for
the analysis of the data presented in Fig.\ 2 of Ref.\
\onlinecite{Takahashi02}. Even the data obtained in
$H=30\,{\text{kOe}}$ (Fig.\ 1 of Ref.\ \onlinecite{Takahashi02})
may not be sufficiently below $H_{c2}^{\|c}$ to yield a proper
$\gamma$ value from a fit of Eq.\ (\ref{tau_rev}) as there can
still be a certain degree of core overlap.

\begin{figure}[tb]
\includegraphics[width=0.95\linewidth]{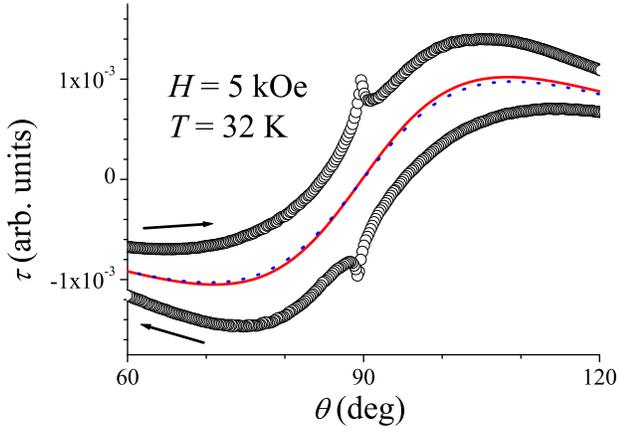}
\caption{Torque $\tau$ of a MgB$_2$ single
crystal\cite{note_sampleA} vs angle $\theta$ at $32 \, {\text{K}}$
in $5\,{\text{kOe}}$. Shown are the measured data ($\circ$), a fit
of Eq.\ (\ref{tau_rev}) to
$(\tau_{\text{inc}}+\tau_{\text{dec}})/2$, in the range of angles
as shown (full line), and a fit of Eq.\ (\ref{tau_rev}) to the
shaked torque (broken line).} \label{fig1}
\end{figure}
\begin{figure}[!tb]
\includegraphics[width=0.95\linewidth]{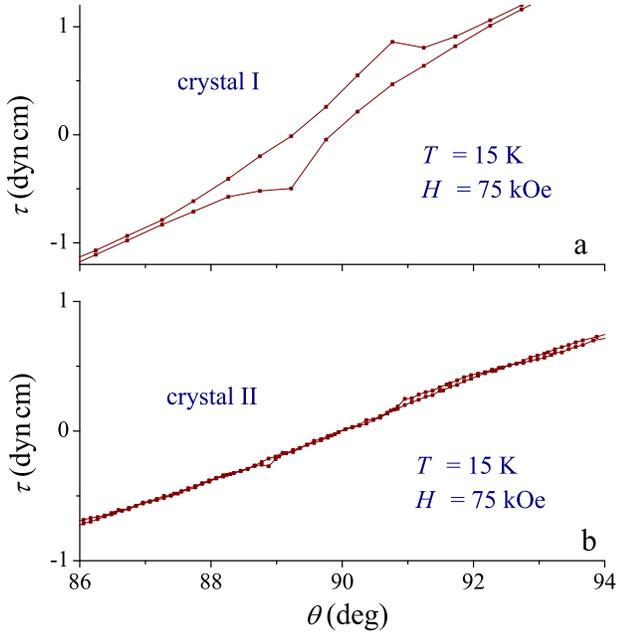}
\caption{Torque $\tau$ vs angle $\theta$ near $H\|ab$ for two
different MgB$_2$ crystals, one showing a peak in the irreversible
torque near $H\|ab$ (a), the other one not showing such a peak
(b).} \label{fig2}
\end{figure}

The data measured in $10\,{\text{kOe}}$ indicate a lower
anisotropy. As Takahashi {\em et al.} correctly note, simple
arithmetic averaging of the torque measured for clockwise
($\tau_{\text{inc}}$) and counterclockwise ($\tau_{\text{dec}}$)
change of the field direction not necessarily yields the correct
equilibrium reversible torque when there is substantial
irreversibility. However, in a torque study on
YBa$_2$Cu$_3$O$_{7-\delta}$, Willemin {\em et
al.}\cite{Willemin98} observed that $\gamma$ extracted from the
correct reversible torque, obtained with ``shaking'' the vortices
by a small perpendicular ac field,\cite{note_shak} had a lower
value than $\gamma$ extracted from
$(\tau_{\text{inc}}+\tau_{\text{dec}})/2$. Thus, substantial
irreversibility leads to an overestimation of the anisotropy
$\gamma$. There is no hard proof that it has to be like this
always. To check the situation in MgB$_2$, we performed a similar
experiment. Figure \ref{fig1} shows $\tau(\theta)$ data measured
on a MgB$_2$ single crystal\cite{note_sampleA} in
$5\,{\text{kOe}}$ at $32\,{\text{K}}$ and the corresponding fit of
Eq.\ (\ref{tau_rev}) to $(\tau_{\text{inc}}+\tau_{\text{dec}})/2$.
The reversible torque data obtained by ``shaking'' can be seen in
Fig.\ 5 of Ref.\ \onlinecite{Angst02MgB2anis}.\cite{note_mistake}
In this case of data obtained with a ``shaking'' experiment, an
anisotropy of $\gamma\simeq 3.2$ was fitted, together with $\eta
H_{c2}^{\|c}/H \simeq 1.75$. When the ``unshaked'' data of Fig.\
\ref{fig1} are fitted, while keeping $\eta H_{c2}^{\|c}/H \simeq
1.75$, $\gamma\simeq 3.4$ is obtained, i.e., a higher value. If
$\eta H_{c2}^{\|c}/H$ is instead fitted as well, even a higher
anisotropy, $\gamma\simeq 3.6$ is obtained. If the whole angular
range from 0 to $180\,{\text{deg}}$ is fitted, the fitted $\gamma$
is also $3.6$, but to compare with the shaked data it is better to
use the same range of angles. Either way, the values of $\gamma$
obtained from non-shaked data are higher and it seems therefore
reasonable to conclude that the value
$\gamma(10\,{\text{kOe}})\simeq 2.84$ obtained in Ref.\
\onlinecite{Takahashi02} represents an upper limit of the real
anisotropy.

If the description of the experimental data from $30$ to
$60\,{\text{kOe}}$ by Eq.\ (\ref{tau_rev}) is unreliable, the
analysis of the data is thus limited to $10\,{\text{kOe}}$ with
the value of $\gamma$ given in Ref.\ \onlinecite{Takahashi02}
representing an upper limit, and to $20\,{\text{kOe}}$. In this
case, we can conclude that the data presented in Ref.\
\onlinecite{Takahashi02} indicate a field dependent effective
anisotropy parameter $\gamma$, in accordance with the conclusions
of Ref.\ \onlinecite{Angst02MgB2anis}.

Takahashi {\em et al.} observed a peak in the irreversible torque
at $10\,{\text{K}}$ in $10\,{\text{kOe}}$ for field directions
close to the $ab$-plane. They interpret the existence of such a
peak as evidence of intrinsic pinning (lock-in effect) and thus as
experimental evidence that superconductivity occurs in the boron
layers, in analogy to the layered cuprate HTSC. The existence of a
similar peak was observed on another single crystal of MgB$_2$
(see Fig.\ 5 of Ref.\ \onlinecite{Angst02MgB2anis}) and also
attributed to the lock-in effect, although no further conclusions
were drawn. However, from subsequent studies of different single
crystals of MgB$_2$, we found that such a peak is present in some
MgB$_2$ crystals, but not in all: Figure \ref{fig2} presents
$\tau(\theta)$ near $H\|ab$ for two different crystals, one
showing a large lock-in effect, and the other one practically
none. On the base of this observation, we conclude that the
presence of the peak is not an intrinsic property of MgB$_2$. This
is in accordance with the estimation that the ratio of the
coherence length perpendicular to the layers $\xi_{c}$ over the
layer separation $d$ (estimated by the $c$-axis of the unit cell)
is larger than $1$. A ratio $\xi_c/d>1$ implies that MgB$_2$ is
within the anisotropic 3D limit, and therefore a lock-in effect
and intrinsic pinning should not be observed. The peak for $H\|ab$
observed in some MgB$_2$ crystals may be caused, for example, by a
small amount of stacking faults.\cite{Zhu01} This problem will be
discussed in more detail in one of our next papers. As a
non-intrinsic feature, the observed peak for $H\|ab$ can not prove
the occurrence of superconductivity in the boron layers.
%\section{Experimental}

%\section{Results and Discussion}

%\section{Conclusions}

%\begin{acknowledgments}
%\end{acknowledgments}

%\def\refname{}
%\newcommand{\noopsort}[1]{} \newcommand{\printfirst}[2]{#1}
%  \newcommand{\singleletter}[1]{#1} \newcommand{\switchargs}[2]{#2#1}
%\begin{thebibliography}{36}
%\end{thebibliography}
%\bibliographystyle{prsty}
%\bibliography{MgB2,da_pin,da_th_em,da_th_mi,da_mate,own}

\begin{thebibliography}{13}
\bibitem[{not({\natexlab{c}})}]{note_mistake}
\bibinfo{note}{Note that the angle values in the insets of Fig.\ 5 correspond
  to the angle between $H$ and the $ab$-plane.}
\end{thebibliography}

\newcommand{\noopsort}[1]{} \newcommand{\printfirst}[2]{#1}
  \newcommand{\singleletter}[1]{#1} \newcommand{\switchargs}[2]{#2#1}

\end{document}